\documentclass[letterpaper,prl,floatfix,titlepage,twocolumn,aps,10pt]{revtex4-1}
\usepackage{
	amssymb, % Math symbols such as \mathbb
	amsthm, % Theorem Formatting
	amsmath,
	mathtools,%this lets us define paired delimiters
	MnSymbol, %for \rrangle and \llangle
	physics,
	bm,
	ifthen,
	etoolbox, %for robust if/then control
	color, % Change text and page color
	appendix,
	phonetic,
	todonotes
} 

\newcommand{\draftmode}{1}    %to control draft colors below
\newcommand{\notetoself}[1]{\ifnum \draftmode=1 \todo[inline,  backgroundcolor=blue!20!white]{#1} \fi}
\newcommand{\cuttext}[1]{\ifnum \draftmode=1 \todo[inline,bordercolor=black!5!white,backgroundcolor=black!5!white]{\color{black!70!white} #1} \fi}
\newcommand{\todoj}[1]{\ifnum \draftmode=1 \todo[inline,backgroundcolor=green!15!white]{#1} \fi}
\newcommand{\warntext}[1]{\ifnum \draftmode=1 \todo[inline, bordercolor=orange!30!white, backgroundcolor=orange!30!white]{#1} \else #1 \fi}  
\let \projector \dyad % in package "physics", dyad{x}{y} makes |x><y| while dyad{x} makes |x><x|.  This just lets us call the second thing "\projector" too.

\DeclarePairedDelimiter\avg{\langle}{\rangle}%  <x>
\DeclarePairedDelimiter\aavg{\llangle}{\rrangle}%   <<x>>
\makeatletter
\let\oldavg\avg
\def\avg{\@ifstar{\oldavg}{\oldavg*}}
\let\oldaavg\aavg
\def\aavg{\@ifstar{\oldaavg}{\oldaavg*}}
\makeatother

\newcommand{\Nor}{\ensuremath{\mathcal{N}}} %\mathcal{P}, \mathcal{N} \mathcal{A}.  
\newcommand{\Nk}{\ensuremath{N}} %P,N, A

\newcommand{\Eth}{\ensuremath{\mathcal{E}}}
\newcommand{\Ek}{\ensuremath{E}}
\newcommand{\mrL}{\ensuremath{\mathrm{1}}}%L
\newcommand{\mrR}{\ensuremath{\mathrm{2}}}%R
\newcommand{\LorR}{\ensuremath{j}}%X
\newcommand{\Dmat}{D}%\bar{D}} %{D}%
\newcommand{\Ddiff}{D}%D^{\mathrm{diff}}} %diffusion coefficent
\newcommand{\Hmat}{H} %Hamiltonian mat
\newcommand{\Fmat}{K} %{\bar{F}} %Forcing mat
\usepackage{phonetic}
\pdfoutput=1

\renewcommand{\draftmode}{0} %1 for drafts (faded text, etc.), 0 to not
\begin{document}
%%%%%%%%%%%%%%%%%%%%%%%%%%%%%

%%%%%%%%%%%%%%%%%%%%%%%%%%%%%
%      Title/authors        %
%%%%%%%%%%%%%%%%%%%%%%%%%%%%%

\title{Decoherence from classically undetectable sources: A standard quantum limit for diffusion}
\date{\today}
\author{C.~Jess~Riedel}\
\affiliation{Perimeter Institute for Theoretical Physics, Waterloo, Ontario N2L 2Y5, Canada}
\affiliation{IBM Watson Research Center, Yorktown Heights, NY, USA}

%%%%%%%%%%%%%%%%%%%%%%%%%%%%%
%         Abstract          %
%%%%%%%%%%%%%%%%%%%%%%%%%%%%%

\begin{abstract}
In the pursuit of speculative new particles, forces, and dimensions with vanishingly small influence on normal matter, understanding the ultimate physical limits of experimental sensitivity is essential.
Here, I show that quantum decoherence offers a window into otherwise inaccessible realms.  There is a standard quantum limit for diffusion that restricts some entanglement-generating phenomena, like soft collisions with new particle species, from having appreciable classical influence on normal matter. Such phenomena are \emph{classically undetectable} but can be revealed by the anomalous decoherence they induce on non-classical superpositions with long-range coherence in phase space. This gives strong, novel motivation for the construction of matter interferometers and other experimental sources of large superpositions, which recently have seen rapid progress.  Decoherence is always at least second order in the coupling strength, so such searches are best suited for \emph{soft} interactions (e.g., small momentum transfers), but not \emph{weak} ones (i.e., small coupling constants).
\end{abstract}

\maketitle

%%%%%%%%%%%%%%%%%%%%%%%%%%%%%
%      Introduction         %
%%%%%%%%%%%%%%%%%%%%%%%%%%%%%

%%%%%      MZ Figure       %%%%%%%%%%%%%%%%%%%%
\begin{figure} [b!]
	\centering 
	\newcommand{\pbwidthfactor}{0.95}
	\includegraphics[width=\pbwidthfactor\columnwidth]{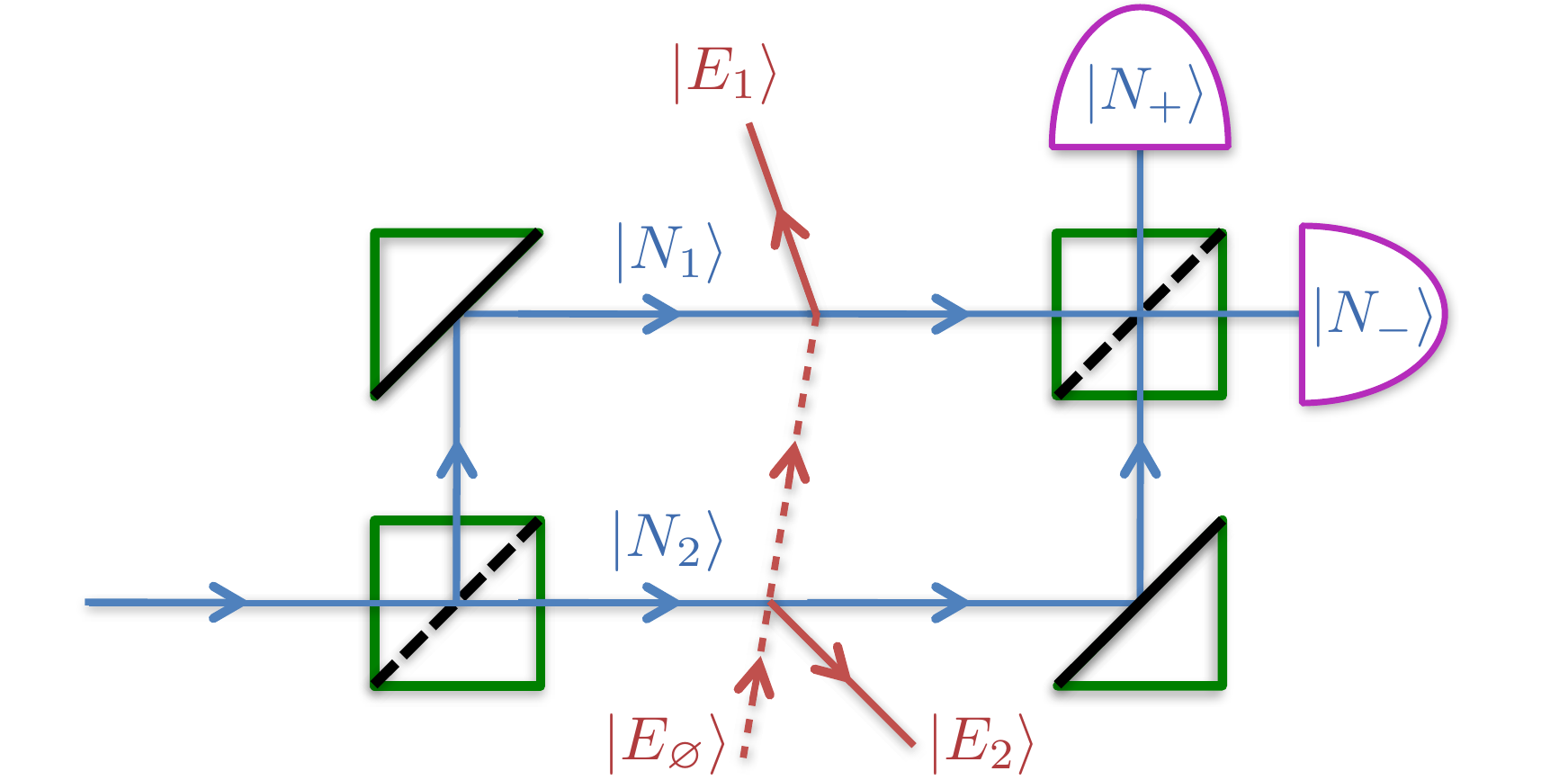}
	\caption{\textbf{Decoherence detection with a Mach-Zehnder interferometer.} A probe initially in a superposition $\ket{\Nk_{1}}+\ket{\Nk_{2}}$ of two wavepackets is effectively measured by an interferometer in the basis $\{\ket{\Nk_{\pm}}=\ket{\Nk_{1}}\pm\ket{\Nk_{2}}\}$.  If the probe scatters a hypothetical particle $\ket{\Ek_\varnothing}$ into distinct conditional states $\ket{\Ek_1}$ and $\ket{\Ek_2}$ while being negligibly deflected, then the paths decohere and the particle will be found in state $\ket{\Nk_{-}}$ with probability $P_- = (1-\mathrm{Re}\braket{\Ek_1}{\Ek_2})/2$.}
	
	\label{fig:mz_diagram}
\end{figure}
%%%%%%%%%%%%%%%%%%%%

Ultrasensitive detectors of feeble signals confront fundamental quantum barriers that can only be beaten using non-classical states of matter and radiation \cite{giovannetti2004quantum-enhanced, adhikari2014gravitational}. One way to capture the sense in which devices exploiting quantum superpositions can be strictly more sensitive than those using classical techniques is to derive a \emph{standard quantum limit} \cite{braginsky1975quantum-mechanical, braginsky1980quantum, braginsky1995quantum,giovannetti2004quantum-enhanced} (SQL).  In general, an SQL is a sensitivity limit arising from the assumption that the experimental probe used to investigate some phenomena can only be prepared and measured in certain ways --- usually corresponding to the position basis --- rather than the vast menu of quantum states and Hermitian observables.  For instance, the most common SQL restricts the detection of weak forces through sequential position measurements of a test mass, a limit that can be beaten by measuring a relative phase shift induced on a coherent spatial superposition of the mass.  This limit now challenges state-of-the-art gravitational wave detectors like LIGO \cite{adhikari2014gravitational}. 

In this article, I derive an analogous SQL for detecting hypothetical weak sources of momentum diffusion, such as a soft flux of a new particle species \cite{joos1985emergence,schlosshauer2008decoherence,hornberger2003collisional,riedel2013direct}, and show that this limit can be beaten using coherent superpositions that decohere rather than acquire a phase shift. I point out that such decoherence only enters at second order in the coupling between the source and the probe, in contrast to the measurement of unitary effects (like gravitational waves) which may be first order.  Finally, I give a precise sense in which hypothetical phenomena, including both forces and diffusion, can be classically undetectable in an $\hbar \to 0$ limit while their influence on quantum experiments remains finite.

Understanding the detective capabilities of experiments producing non-classical superpositions is a timely topic; bigger superpositions (as measured by mass, spatial separation, and lifetime) are more sensitive to decoherence, and there has been spectacular recent successes in generating and verifying such states \cite{arndt2014testing}.  
%broader perspective
Large molecules containing over ten thousand nucleons have been interfered in diffractive-slit experiments \cite{eibenberger2013matterwave}, with orders of magnitude of improvement predicted in the medium-term future \cite{hornberger2012colloquium,haslinger2013universal,arndt2014testing}.  Optically trapped nanoparticles promise to achieve wide spatial superpositions of even greater amounts of matter \cite{romero-isart2011large,bateman2014near-field}---perhaps reaching an astounding $10^{10}$ amu in spaceborne experiments \cite{kaltenbaek2013testing,kaltenbaek2015macroscopic}---and quantum micromechanical resonators will involve over $10^{14}$ amu, albeit with femtometer superposition separations \cite{oconnell2010quantum,aspelmeyer2014cavity}.  Such massive  superpositions are predicated on the careful suppression of conventional sources of decoherence, making them unusually sensitive to anomalous decoherence from novel sources.  General strategies exist to distinguish novel sources of decoherence from merely misunderstood conventional ones \cite{riedel2013direct}.

\emph{Framework.}---Consider an interferometer (Fig.~\ref{fig:mz_diagram}) producing a coherent superposition $\ket{\Nk_1} + \ket{\Nk_2}$ of some probe (e.g., a test mass), where $\ket{\Nk_1}$ and $\ket{\Nk_2}$ are wavepackets separated by a large distance $L$.  The wavepackets are recombined and, by way of appropriately aligned mirrors and detectors, the probe is effectively measured in the basis $\{\ket{\Nk_\pm} = \ket{\Nk_1} \pm \ket{\Nk_2}\}$.  Up to experimental error, the result is always $\ket{\Nk_+}$.  Now imagine that during its flight through the interferometer the probe interacts with some other hypothetical quantum system in an initial state $\ket{\Ek_{\varnothing}}$ such that the hypothetical system is disturbed but there is negligible influence on either probe wavepacket individually.  That is, assume that the evolution is well approximated by the form 

\begin{align}
\big[\ket{\Nk_1} + \ket{\Nk_2}\big] \ket{\Ek_{\varnothing}} \to \ket{\Nk_1} \ket{\Ek_1} + \ket{\Nk_2} \ket{\Ek_2}
\end{align}
where $\ket{\Ek_1}$ and $\ket{\Ek_2}$ are arbitrary conditional states of the hypothetical system. 

If $\ket{\Ek_1}$ and $\ket{\Ek_2}$ differ only by a phase, $\braket{\Ek_1}{\Ek_2} = e^{i \theta}$,  then the probe remains in a pure state and, were it observed before the wavepackets are recombined at the end of the interferometer, would still be found with equal likelihood in either arm.  Nevertheless, the relative phase between the two wavepackets can be inferred from the fact that the effective measurement at the end of the interferometer will now result in outcome $\ket{\Nk_-}$ with probability $P_- = (1-\cos \theta)/2$.  This is in fact the basic mechanism by which atom interferometers are proposed to detect very weak forces like gravitational waves \cite{dimopoulos2009gravitational}, which are too small to displace atoms by a distance comparable to their wavepacket size yet still leave a measurable relative phase.

But it's also possible to consider scenarios for which $\braket{\Ek_1}{\Ek_2} \approx 0$, such as flux of soft particles scattering into distinguishable out states \cite{joos1985emergence,schlosshauer2008decoherence,hornberger2003collisional,riedel2013direct}.  In this case, the probe is decohered and the probabilities for both outcomes of the measurement are equal: $P_+ = P_- = 1/2$.  Just as for relative phases induced by weak forces, this decoherence can be detected even though the \emph{classical} effects on the probe wavepackets (spatial displacements or momentum transfers) are completely negligible.

%%%      Wig Figure       %%%%%%%%%%%%%%%%%%%%
\begin{figure} [b!]
	\centering 
	\newcommand{\pbwidthfactor}{0.95}
	\includegraphics[width=\columnwidth]{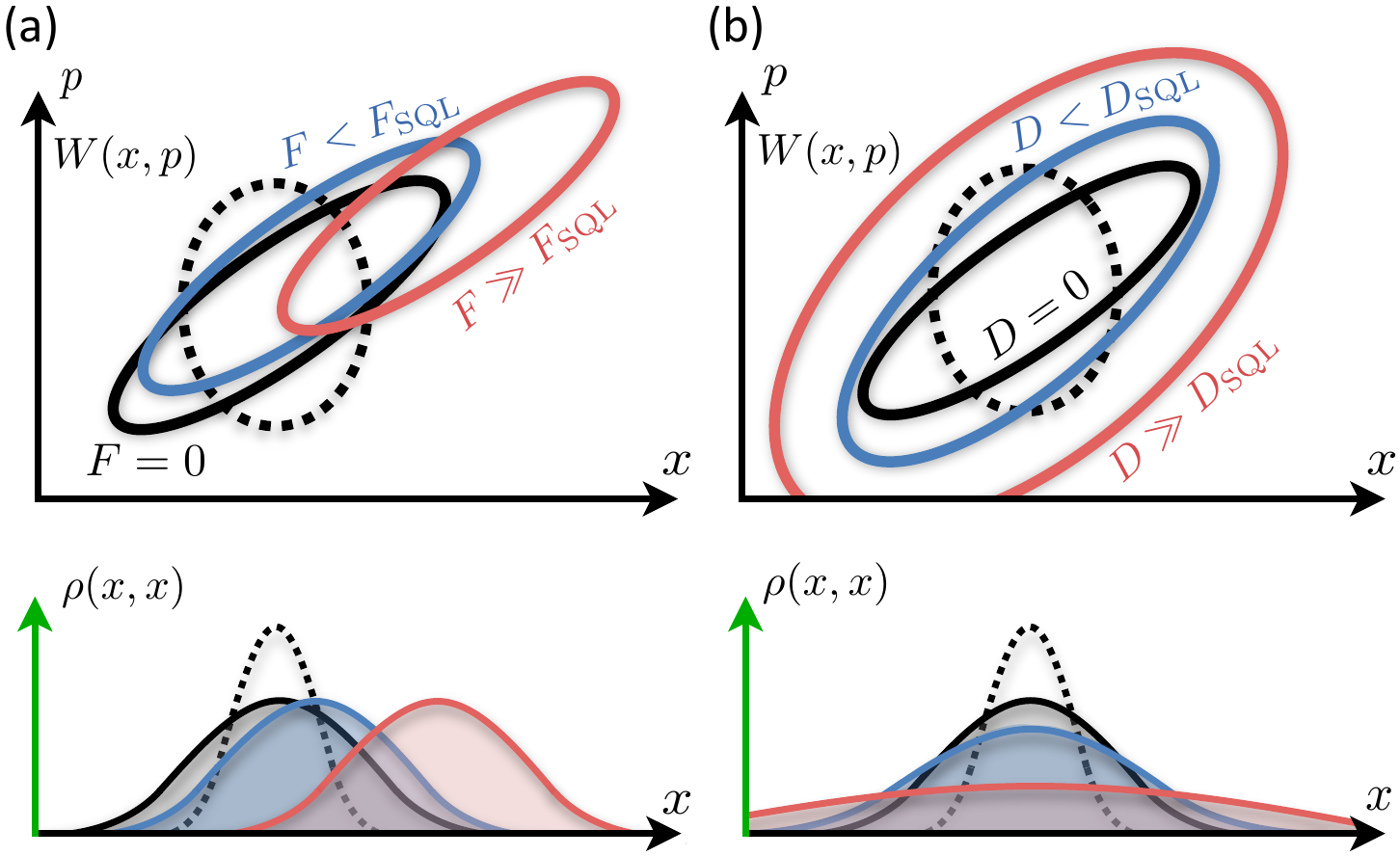}
	\caption{\textbf{Standard quantum limit for forces and momentum diffusion.} A test mass is initially placed in a minimal uncertainty wavepacket with a Wigner distribution $W(x,p)$ over phase space (top) that contains the bulk of its mass within a 2$\sigma$-contour of a Gaussian distribution (dashed black line).  After a time $T$, the state will spread in position due to wavefunction dispersion even in the absence of a force or diffusion (solid black).  The probability distribution of a subsequent position measurement (bottom) is obtained by integrating the Wigner distribution over momentum.  (a) If a uniform force $F$ displaces the wavepacket, it will not be reliably distinguishable from the zero-force scenario when $F < F_{\mathrm{SQL}}$ (blue), in contrast to the case $F \gg F_{\mathrm{SQL}}$ (red). (b) The standard quantum limit for momentum diffusion, $\Ddiff_{\mathrm{SQL}}$, provides a similar threshold for the detection of an environment that weakly smears the quantum state in phase space.}
	\label{fig:wig_diagram}
\end{figure}

%%%%%%%%%%%%%%%%%%%
Although searches for anomalous decoherence arising from speculative theories of quantum gravity \cite{percival1997detection} and modifications of quantum mechanics \cite{nimmrichter2011testing} have been considered before, it is only recently that this technique has been proposed as a way to detect new particles within an otherwise conventional framework.  In particular, certain classes of low-mass (sub-MeV) dark matter --- which would be invisible to traditional direct-detection experiments --- could be identified by the characteristic decoherence they induce in future matter interferometers \cite{riedel2013direct}.

%%      Wig tilt Figure       %%%%%%%%%%%%%%%%%%%%
\begin{figure*} [!] 
	\centering 
	\newcommand{\pbwidthfactor}{0.95}
	\includegraphics[width=\textwidth]{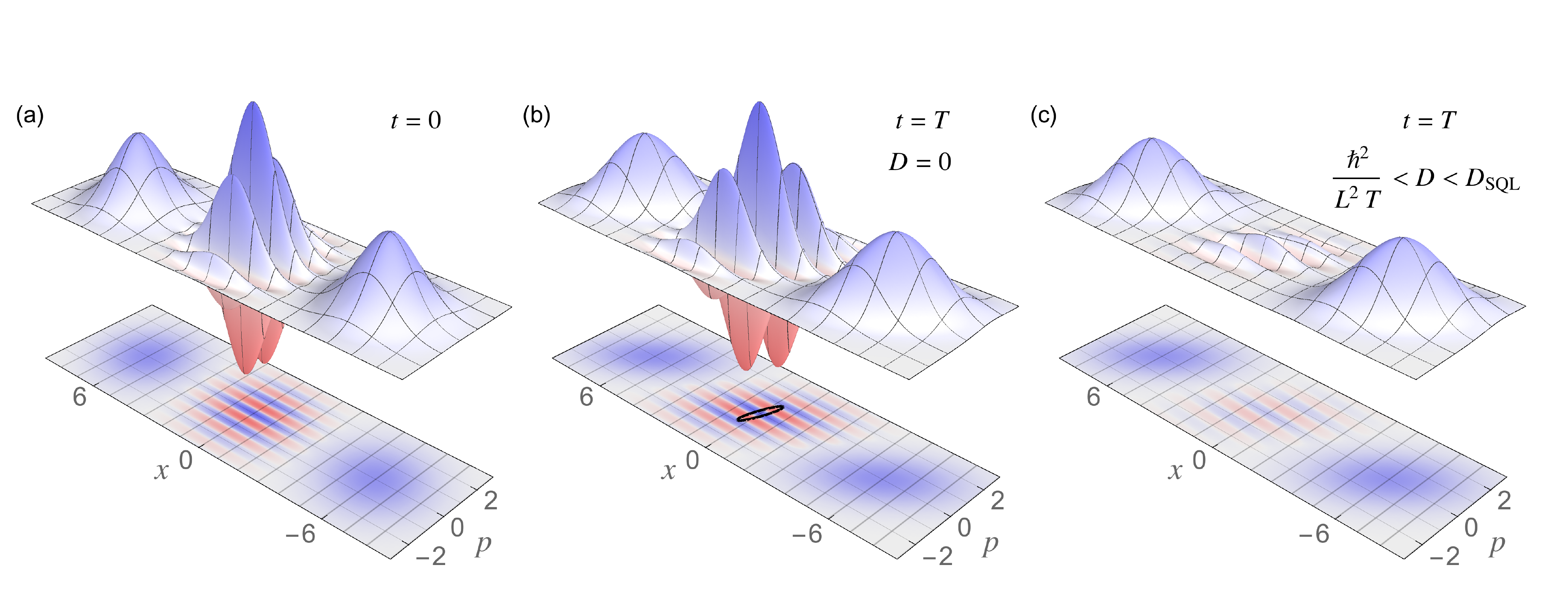}
	\caption{\textbf{Diffusion sensitivity from fine phase-space oscillations.} (a) The Wigner distribution over phase space for a coherent superposition of two minimal uncertainty wavepackets for a particle free to move in one dimension.  The wavepackets have position spread $\sigma_x^{\mathrm{prep}}=\sqrt{\hbar T/m}$, average position $x_0 = \pm 6\sigma_x^{\mathrm{prep}}$ (so $L = 2x_0 = 12 \sigma_x^{\mathrm{prep}}$), and average momentum $p_0 = 0$.  The grid units are $\sigma_x^{\mathrm{prep}}$ and $\sigma_p^{\mathrm{prep}}=\hbar/2(\sigma_x^{\mathrm{prep}})$. The fine (sub-$\hbar$) phase-space oscillations near the origin are indicative of coherence \cite{toscano2006sub-planck}.  (These oscillations are absent from an incoherent mixture of wavepackets, for which the Wigner distribution is given by the simple sum of the wavepackets.)  (b) Under free evolution, the position and momentum of the particle become positively correlated, skewing the distribution.  The additional effects of momentum diffusion over the time period can then be obtained by convolving this freely-evolved Wigner distribution with a Gaussian kernel \cite{brodier2004symplectic, diosi2002exact}; the 5$\sigma$-contour of this kernel for one choice of diffusion strength below the SQL is given by the black oval.  This kernel is large compared to the oscillations, but small compared to the wavepacket widths. (c) The Wigner distribution after the initial state is evolved in the presence of momentum diffusion.  The smoothing induced by the convolution only negligibly distorts the wavepackets, but the fine oscillations are destroyed, indicating that the state is almost fully decohered.  Without the quantum coherence there would be no oscillations and hence very little sensitivity to the diffusion.  The wavenumber of the oscillations is proportional to the separation $L$, so arbitrarily weak diffusion over a fixed time can be detected for sufficiently large separation.}
	\label{fig:wig_tilt}
\end{figure*}
%%%%%%%%%%%%%%%%%%%

\emph{Standard quantum limit.}---Let us recall the popular heuristic argument for the SQL for the detection of weak forces \cite{braginsky1975quantum-mechanical}.   A single quantum degree of freedom $x$, which we will imagine to be the position of a test mass $m$ along one dimension, evolves in the presence of a small, spatially uniform force $F$ for some time interval $T$.  At $t=0$, the mass is prepared by a position measurement of finite precision in a spatially localized pure state with some finite, adjustable width $\sigma_x^{\mathrm{prep}}$.  By the uncertainty principle, this prepared state necessarily has momentum width at least as large as $\sigma_p^{\mathrm{prep}} = \hbar/(2 \sigma_x^{\mathrm{prep}})$.  During the time period $T$ the momentum width produces an additional position spread $\sigma_x^{\mathrm{disp}} =  \int \dd{T} \sigma_p^{\mathrm{prep}} /m = \hbar T / (2 m \sigma_x^{\mathrm{prep}})$ through wavefunction dispersion.  The total uncertainty for a final position measurement is then obtained by adding in quadrature: $(\sigma_x^{\mathrm{meas}})^2 = (\sigma_x^{\mathrm{prep}})^2 + (\sigma_x^{\mathrm{disp}})^2$. A force translates the wavepacket uniformly \cite{combescure2012coherent}  by $\Delta x_F = F T^2/(2m)$ relative to the null ($F=0$) dynamics, and so the force will not be detectable if this distance is much smaller than $\sigma_x^{\mathrm{meas}}$.  One cannot simply make the initial position uncertainty $\sigma_x^{\mathrm{prep}}$ arbitrarily small to increase sensitivity because of the resulting larger wavepacket dispersion  $\sigma_x^{\mathrm{disp}}$.  Minimizing $\sigma_x^{\mathrm{meas}}$ with respect to the choice of initial width $\sigma_x^{\mathrm{prep}}$ yields $\sigma_x^{\mathrm{meas}} = \sqrt{2} \sigma_x^{\mathrm{prep}} = \sqrt{\hbar T/m}$.  We then obtain the requirement that a force must satisfy $F \gtrsim F_{\mathrm{SQL}}$ to be detectable, where $F_{\mathrm{SQL}} \equiv 2 \sqrt{\hbar m / T^3}$ is the SQL for weak-force detection with mass $m$ acted upon for time interval $T$.  See Fig.~\ref{fig:wig_diagram}~(a).

Now we generalize to an open quantum system, allowing the mass to interact with an environment so that its reduced dynamics are generally not unitary.  We seek to understand the physical limits for detecting the effects of an environment that weakly decoheres the test mass.  For ideal quantum Brownian motion (QBM) in the frictionless limit, the density matrix of the test mass is governed by the master equation \cite{joos1985emergence, unruh1989reduction, isar1994open, brodier2004symplectic}
\begin{align}
\label{eq:decoh}
\partial_t \rho_t = - \frac{i}{\hbar} [\hat{H},\rho_t] - \frac{\Ddiff}{\hbar^2} [\hat{x},[\hat{x},\rho_t]]
\end{align}
where $\hat{H}=\hat{p}^2/2m$ and where $\Ddiff$ is the momentum diffusion coefficient.  One can get some intuition for the resulting decoherence by noting \cite{joos1985emergence, isar1994open, brodier2004symplectic} that the off-diagonal terms of the position-basis density matrix are exponentially suppressed for sufficiently large separation $x-x'$: $\rho_t(x,x') \approx \rho_0(x,x') \exp[-\Ddiff t (x-x')^2/\hbar^2]$.

Equation \eqref{eq:decoh} is the simplest possible model of spatial decoherence \footnote{See the Appendix for an overview of ideal QBM.}, playing the pedagogical role of a harmonic oscillator for the study of open quantum systems.  It is widely applicable due to the ubiquity of spatially local interactions, providing a good approximation to the effects of a thermal bath of linearly coupled oscillators \cite{unruh1989reduction} (although note limitations \cite{anglin1997deconstructing, ford2005limitations}).  Importantly, \eqref{eq:decoh} well describes the dynamics taken by a test mass subjected to \emph{collisional decoherence} \cite{joos1985emergence, gallis1990environmental, schlosshauer2008decoherence,diosi2002exact} from an environment of lighter particles, such as a gas \cite{hornberger2006master}, blackbody radiation \cite{joos1985emergence}, or low-mass dark matter \cite{riedel2013direct}.  For a given environment, the approximation becomes more accurate for larger test masses.

Intuitively, diffusion $\Ddiff$ that is small compared to other parameters will be negligible, so like the force SQL ($F_{\mathrm{SQL}} \sim \sqrt{\hbar m / T^3}$) we can guess the diffusion SQL up to a numerical factor by dimensional analysis: $\Ddiff_{\mathrm{SQL}} \sim m \hbar/ T^2$.  This can be fleshed out with a heuristic argument analogous to the one above \footnote{There is a well-known loophole in the heuristic argument for the force SQL arising when the probe is prepared in the so-called contractive states \cite{yuen1983contractive,lynch1984comment,yuen1984yuen,caves1985defense,ozawa1988measurement}.  This also applies to the diffusion SQL, and is addressed in the Appendix.}. Let the initial state of the test mass be a wavepacket with spatial width $\sigma_x^{\mathrm{prep}}=\sqrt{\hbar T/(2m)}$ minimizing (in the absence of diffusion) $\sigma_x^{\mathrm{meas}}$.  The momentum diffusion induced by \eqref{eq:decoh} causes the test mass to follow a random walk in momentum space: $(\sigma_p^{\mathrm{diff}})^2 = 2 \Ddiff T$, yielding $\sigma_x^{\mathrm{diff}} = \int\! \dd T \, \sigma_p^{\mathrm{diff}}/m =  \sqrt{8 D T^3}/(3m)$.  The diffusion can only be reliably detected when $\sigma_x^{\mathrm{diff}} \gtrsim \sigma_x^{\mathrm{meas}}$.  Therefore, the diffusion SQL is that the diffusion strength obey $\Ddiff \gtrsim \Ddiff_{\mathrm{SQL}}$ for $\Ddiff_{\mathrm{SQL}} \equiv 9 \hbar m/(8 T^2)$. When $\Ddiff \ll \Ddiff_{\mathrm{SQL}}$, the diffusion cannot be detected.  See Fig.~\ref{fig:wig_diagram}~(b), and the Appendix for a rigorous derivation.

Like the force SQL, the diffusion SQL can be beaten by placing the test mass in a coherent superposition of two wavepackets widely separated in phase space (Fig.~\ref{fig:wig_tilt}).  To see this, consider an initial wavepacket $\vert \Nk_1 \rangle$ that minimizes $\sigma_x^{\mathrm{meas}}$.  If $\Ddiff \ll \Ddiff_{\mathrm{SQL}}$ then such a wavepacket is left essentially undisturbed by the diffusion over the timescale $T$.  However, if a coherent superposition of $\vert \Nk_1 \rangle$ with $\vert \Nk_2 \rangle$ is prepared, where the latter is simply the former translated by a sufficiently long distance $L \gg \hbar/\sqrt{\Ddiff T}$, then the superposition decoheres into an incoherent mixture of the two wavepackets within a decoherence time $\tau_\Ddiff= \hbar^2/(\Ddiff L^2) \ll T$.  The wideness of the superposition, rather than its mere existence, is important. The smallest diffusion coefficient that effectively decoheres a superposition of extent $L$ is about $\Ddiff_{\mathrm{min}} \equiv \hbar^2/(T L^2)$, which reduces to the classical sensitivity $\Ddiff_{\mathrm{SQL}}$ when the superposed wavepackets are minimally separated, i.e., when the separation $L$ is of order the optimal wavepacket size $\sigma_x^{\mathrm{meas}} = \sqrt{\hbar T/m}$. 

\emph{No entanglement at first order.}---It is worth noting that a probe $\Nor$ cannot become entangled with another system $\Eth$ at first order in the Hamiltonian that couples them, suggesting that anomalous decoherence is a method best suited for detecting interactions that are very \emph{soft} (e.g., small momentum transfers) but not \emph{weak} (i.e., small coupling constant).

%Here, ``soft'' means that the classical effects, such as momentum transfer and spatial displacement, are small compared to relevant scales in the interaction, as for soft photons.  ``Weak'' means that the coupling constant between $\Nor$ and $\Eth$ is small.  (For example, neutrino collisions through the weak force are very rare on account of the smallness of the Fermi coupling constant, but the typical momentum transfer in any collision event is often large.)

Let $\Nor$ and $\Eth$ initially be in an uncorrelated pure state, $\vert \psi_0 \rangle = \vert \Nk_0 \rangle \otimes \vert \Ek_0 \rangle$, and suppose they evolve under $U_t = \exp(- i H t)$ where the Hamiltonian is of the form $H=H_\Nor + H_\Eth + \epsilon H_I$.  Here, $H_\Nor = H_\Nor \otimes I_\Eth$ and $H_\Eth = I_\Nor \otimes H_\Eth$ are the self-Hamiltonians for $\Nor$ and $\Eth$, respectively, and $\epsilon H_I$ is the interaction. Local unitaries cannot change the entanglement, so without loss of generality we can consider the modification $U'_t = e^{+i H_\Nor t} e^{+i H_\Eth t} e^{-i H t}$, which peels off the unimportant local evolution of $\Nor$ and $\Eth$.  We will use the Zassenhaus formula \cite{casas2012efficient}
\begin{align}
e^{A+B} = e^A e^B \prod_{n=2}^{\infty} e^{C_n(A,B)},
\end{align}
where $C_n(A,B)$ is a homogeneous Lie polynomial of degree $n$ with, for example, 
%\begin{gather}
%C_2(A,B) = - \frac{1}{2} [A,B], \\
%C_3(A,B) = \frac{1}{3} [B,[A,B]] + \frac{1}{6} [A,[A,B]].
%\end{gather}
$C_2(A,B) = - \frac{1}{2} [A,B]$ and $C_3(A,B) = \frac{1}{3} [B,[A,B]] + \frac{1}{6} [A,[A,B]]$.
Then, using $A=-it(H_\Nor + H_\Eth)$, $B=-i t \epsilon H_I$, and the fact that $H_\Nor$ and $H_\Eth$ commute, we get
\begin{align}
U'_t = e^{-i \epsilon \tilde{H}_I t} \big[1 + O(\epsilon^2) \big]
\end{align}
for some new $\tilde{H}_I$ that is independent of $\epsilon$, because each $C_n(-it(H_\Nor + H_\Eth),-it\epsilon H_I)$ is at least first order in $\epsilon$.  So the probe's density matrix is likewise pure to this accuracy:
\begin{align}
\rho_\Nor(t) = \mathrm{Tr}_\Eth \left[ U'_t \vert \psi_0 \rangle \langle \psi_0 \vert U_t^{\prime \dagger} \right] =  \vert \tilde{\Nk}_t \rangle \langle \tilde{\Nk}_t \vert + O(\epsilon^2)
\end{align}
where $\vert \tilde{\Nk}_t \rangle = \left( I - i \epsilon t \langle \Ek_0  \vert \tilde{H}_I \vert  \Ek_0 \rangle \right) \vert \Nk_0 \rangle$.

Decoherence of an initially pure state necessarily requires entanglement with an environment, so a signal of anomalous decoherence, which is linear in the off-diagonal elements of $\rho_\Nor(t)$, would be at least second order in the coupling.  Therefore, interactions with very tiny coupling constants (e.g., the Fermi constant in searches for relic neutrinos) are poor candidates for detection through decoherence.  Rather, an anomalous decoherence search is a method best suited for interactions with large coupling but with classical effects (e.g., momentum transfers or spatial displacements) that are negligible.   Contrast this with traditional quantum enhanced measurements of phase shifts, such as from weak gravitational waves, which are first order in the coupling strength (although they may still be small relative to decoherence for other reasons \cite{riedel2013direct}).

\emph{Classical undetectability.}---The exceptional sensitivity of superpositions to both displacement and decoherence can be attributed \cite{zurek2001sub-planck, toscano2006sub-planck} to fine structure in a test mass's Wigner distribution at scales smaller than $\hbar$.  Indeed, the decoherence of a superposition of widely separated wavepackets is the ``complexification'' of the phase shift induced by a force.  Let us consider the class of completely positive (CP) reduced dynamics $\Phi$ for the probe's density matrix that leave two wavepackets undisturbed, as in Fig.~\ref{fig:mz_diagram}: $\Phi[\projector{\Nk_\LorR} ] \approx \projector{\Nk_\LorR}$ for $\LorR = \mrL, \mrR$. One can check that any CP map satisfying this requirement must operate on the two-dimensional subspace spanned by the (essentially orthogonal) wavepacket states as
\begin{align}
\Phi[\rho] = \Phi \left[\left( \begin{array}{cc}\rho_{\mrL \mrL} &  \rho_{\mrL \mrR} \\ \rho_{\mrR \mrL} & \rho_{\mrR \mrR}\end{array} \right) \right]=
\left( \begin{array}{cc}\rho_{\mrL \mrL} & \gamma \rho_{\mrL \mrR} \\ \gamma^*\rho_{\mrR \mrL} & \rho_{\mrR \mrR}\end{array} \right),
\end{align}
where $\gamma = \braket{\Ek_1}{\Ek_2} = e^{-s+i \theta}$ is the \emph{decoherence factor}, with $\theta$ real and $s>0$.

The classical force obeys $\theta = FLT/\hbar$ and (since it is unitary) $s=0$.  Decoherence from momentum diffusion, \eqref{eq:decoh}, obeys $s = \Ddiff L^2 T/ \hbar^2$ and (since it is isotropic) $\theta = 0$.  These can be obtained while still satisfying $\Phi[\projector{\Nk_\LorR} ] \approx \projector{\Nk_\LorR}$ to arbitrary accuracy by holding $L$, $T$, $F/\hbar$, and $\Ddiff/\hbar^2$ fixed and taking $\hbar \to 0$, so that $F, \Ddiff \to 0$.  In other words, in the classical limit where wavepackets become points in phase space \cite{zurek1986reduction,zurek1993preferred} (i.e., states of arbitrarily precise position and momentum), the phase and decoherence induced by $\Phi$ in an interferometer can remain fixed --- and hence detectable --- even as the classical strength of the force $F$ and the diffusion $\Ddiff$ go to zero.  Similar limiting behavior exists if the force or diffusion have more complicated dependence on position, or if both effects occur simultaneously  (e.g., the anisotropic dark matter ``wind'' \cite{riedel2013direct}). It is in this sense that coherent superpositions enable the detection of phenomena that are \emph{classically undetectable}.

%%%%%%%%%%%%%%%%%%%%%%%%%%%%%
%       Bibliography        %
%%%%%%%%%%%%%%%%%%%%%%%%%%%%%
\bibliographystyle{apsrev4-1}
\bibliography{zotriedel}

%%%%%%%%%%%%%%%%%%%%%%%%%%%%%

%%%%%%%%%%%%%%%%%%%%%%%%%%%%%%
\section{Acknowledgements}
%%%%%%%%%%%%%%%%%%%%%%%%%%%%%%

I thank Charlie Bennett, Lajos Di\'{o}si, Vittorio Giovannetti, Claus Kiefer, Gordan Krnjaic, Lorenzo Maccone, Igor Pikovski, Graeme Smith, John Smolin, Dan Sank, Mark Srednicki, Shuyi Zhang, Wojciech Zurek, and Michael Zwolak for discussion.  Research at the Perimeter Institute is supported by the Government of Canada through Industry Canada and by the Province of Ontario through the Ministry of Research and Innovation.  This work was also supported by the John Templeton Foundation through grant number 21484.

%%%%%%%%%%%%%%%%%%%%%%%%%%%%%%
\appendix
\section{Ideal QBM and contractive states}
%%%%%%%%%%%%%%%%%%%%%%%%%%%%%%

It is well known that the popular heuristic argument for the standard quantum limit (SQL) for weak forces \cite{braginsky1975quantum-mechanical, braginsky1980quantum, braginsky1995quantum, giovannetti2004quantum-enhanced} contains a loophole \cite{yuen1983contractive,lynch1984comment,yuen1984yuen,caves1985defense,ozawa1988measurement}, and this issue applies likewise to the argument for the diffusion SQL.  Even when restricting to the simple case of a uniform force acting on an initial Gaussian state of the probe that is sensibly described by its first and second moments, a counter-example is obtained by preparing the  probe in a ``contractive'' (but still Gaussian) state for which position and momentum are negatively correlated \cite{yuen1983contractive,ozawa1988measurement}.  (Contractive states can equivalently be defined as squeezed coherent states such that the line in phase space along which the state is positively squeezed has negative slope.)   Like coherent superpositions, contractive states are non-classical and very difficult to prepare experimentally.  A rigorous SQL can be obtained in the general case of ideal quantum Brownian motion (QBM), which encompasses both weak forces and diffusion, when non-contracting Gaussian initial states are assumed.

First let us place our model of momentum diffusion and spatial decoherence in a general framework of ideal QBM.  The simplest type of open quantum system dynamics are Markovian and time-homogeneous, forming a quantum dynamical semigroup for which the purity of the state is nonincreasing with time \cite{lindblad1976generators, alicki2007quantum}.   Setting $\hbar=1$, such evolution is described in generality by a Lindblad master equation
\begin{align}
\label{eq:QBM-general}
\partial_t \rho_t = - i [\hat{H},\rho_t] +\sum_i \left[ \hat{L}^{(i)} \rho_t \hat{L}^{(i) \dagger} - \frac{1}{2} \{ \hat{L}^{(i) \dagger} \hat{L}^{(i)} , \rho_t \} \right].
\end{align}
with the unitary part generated by the Hamiltonian $\hat{H}$ and the non-unitary part by the Lindblad operators $\hat{L}^{(i)}$.  The smallest class of Hamiltonians containing the harmonic oscillator and invariant under rotations and translations in phase space are those that are polynomials no more than quadratic in the phase space variables $x$ and $p$.  Constant terms in Lindblad operators can be absorbed into the Hamiltonian so, to generalize to open quantum systems, we consider the simplest nontrivial Lindblad operators: those linear in $x$ and $p$.  Linear terms in the Hamiltonian can then be eliminated by translating coordinates $x \to x + x_0$, $p \to p + p_0$, putting the origin at the center of the ellipse or hyperbola defined by $\hat{H}$.  (This can't always be done in special cases when quadratic parts of $\hat{H}$ are zero, such as for a uniform force, which must be handled separately  \cite{brodier2004symplectic}.)  Using Einstein summation ($a,b,\ldots = x,p$) over the two phase-space coordinates and letting $\hat{\alpha} = (\hat{x},\hat{p})$, we parameterize the dynamical operators as $\hat{H} = \Hmat_{ab} \hat{\alpha}^a \hat{\alpha}^b/2$ and $\hat{L}^{(i)} = L^{(i)}_{a} \hat{\alpha}^a$, where $\Hmat_{ab}$ is a real symmetric matrix and the $L^{(i)}_{a}$ are complex vectors.  We can then change variables to $\Dmat_{ab} =  \mathrm{Re} \sum_i  (L^{(i)}_{a})^* L^{(i)}_{b}$, $\lambda =   \frac{1}{2i} \epsilon^{ab} \sum_i  (L^{(i)}_{a})^* L^{(i)}_{b} = \mathrm{Im} \sum_i (L^{(i)}_{x})^* L^{(i)}_{p} $, and $\Fmat_{ab} = \Hmat_{ab} + \epsilon_{ab} \lambda$, where $\epsilon^{ab}$ is the antisymmetric Levi-Civita symbol with $\epsilon^{12}=+1=\epsilon_{21}$.   This specializes \eqref{eq:QBM-general}  to QBM:
\begin{align}
\label{eq:QBM-general-alt}
%\partial_t \rho_t = -\frac{i}{2} \Fmat_{ab}  \left[\hat{\alpha}^a,\left\{ \hat{\alpha}^b, \rho_t \right\} \right] - \frac{1}{2}\Dmat_{ab} \left[ \hat{\alpha}^a,\left[\hat{\alpha}^b, \rho_t \right] \right].
\partial_t \rho_t = - \frac{i}{2} \Fmat_{ab}  \left[\hat{\alpha}^a,\left\{ \hat{\alpha}^b, \rho_t \right\} \right] - \frac{1}{2}\Dmat_{ab} \left[ \hat{\alpha}^a,\left[\hat{\alpha}^b, \rho_t \right] \right].
\end{align}

We move to the Wigner representation, where the quantum state $\rho_t$ is exchanged for a quasiprobability distribution $W_t(x,p)$ over phase space: 
\begin{align}
W_t(x,p) &= \int \! \frac{\dd{\Delta x}}{2\pi} \, e^{-i p \Delta x} \rho_t(x + \frac{\Delta x}{2}, x - \frac{\Delta x}{2}) 
\end{align}
such that $\int \! \mathrm{d} x \, W_t(x,p) = \rho_t(p,p)$ and $\int \! \mathrm{d} p \, W_t(x,p) = \rho_t(x,x)$. 
Under \eqref{eq:QBM-general-alt}, the Wigner distribution corresponding to the state $\rho_t$ obeys a Klein-Kramers equation \cite{isar1994open,brodier2004symplectic}
\begin{align}
\label{eq:wig-decoh-general}
%\frac{\partial}{\partial t} W_t (\alpha) = [ \Fmatmod_{ab} \partial_a \alpha_b + \Dmatmod_{ab} \partial_a \partial_b ] W_t(\alpha)
%\frac{\partial}{\partial t} W_t (\alpha) = [ \Fmat^a_{\phantom{a}b} \partial_a \alpha^b + \frac{1}{2}\Dmat^{ab} \partial_a \partial_b ] W_t(\alpha)
\frac{\partial}{\partial t} W_t (\alpha) = [ -\Fmat^a_{\phantom{a}b} \partial_a \alpha^b + \frac{1}{2}\Dmat^{ab} \partial_a \partial_b ] W_t(\alpha)
\end{align}
where $\alpha = (x,p)$, $\partial_a = \partial/\partial \alpha^a$, $\Dmat^{ab} = \epsilon^{ac} \epsilon^{bd} \Dmat_{cd}$, and $\Fmat^a_{\phantom{a}b} = \epsilon^{ac}\Fmat_{cb}$.

The only terms that generate non-Hamiltonian evolution are the real parameter $\lambda = -\Fmat^a_{\phantom{a}a}/2$ and the real symmetric matrix $\Dmat^{ab}$.  The matrix $\Dmat^{a b}$ must \cite{isar1994open} have non-negative eigenvalues and satisfy $\abs{\Dmat^{ab}} \ge \lambda^2$, where $\abs{\Dmat^{ab}} = D_{ab}D^{ab}/2$ is the matrix determinant. One can always diagonalize $\Dmat^{a b}$ through a coordinate rotation in phase space, and the resulting diagonal elements quantify the diffusion in the new $x$ and $p$ directions; by the symplectic symmetry of \eqref{eq:wig-decoh-general}, they have mathematically similar behavior.

Although it is possible to consider general SQLs for measuring all these parameters, they are closely related by symmetries and our primary concern is with the special case of pure momentum diffusion discussed in the main text \eqref{eq:decoh}.  This is when $\Fmat^{x}_{\phantom{x}p}=1/m$, $\Dmat^{pp} = 2 \Ddiff$, and all other coefficients of $\Fmat^{a}_{\phantom{a}b}$ and $\Dmat^{ab}$ are zero. In the Wigner representation, \eqref{eq:wig-decoh-general} reduces to
\begin{align}
\label{eq:wig-decoh}
\frac{\partial}{\partial t} W_t(x,p) = \left[-\frac{p}{ m} \frac{\partial}{\partial x} +  \Ddiff \frac{\partial^2}{\partial p^2} \right] W_t(x,p),
\end{align}
which is identical in form to the Klein-Kramers equation for a classical phase-space probability distribution exhibiting diffusion in momentum space, i.e., ideal classical frictionless Brownian motion \cite{isar1994open}. (This is very different from the large-friction limit studied by Einstein and Smoluchowski.)

The exact solution to \eqref{eq:wig-decoh} for any initial Wigner distribution $W_0(\alpha)$ is \cite{brodier2004symplectic, diosi2002exact} 
\begin{align}
W_t(\alpha) = g_t(\alpha) \star W_0(R_{-t}(\alpha))
\end{align}
where $R_{-t}(\alpha) = R_{-t}(x,p) = (x - p t/m,p)$ is the reversed Hamiltonian evolution, $\star$ denotes the convolution operator in phase space, and $g_t(\alpha) \propto \exp(-(C_t^{-1})^{ab}\alpha_a \alpha_b/2)$ is the normalized Gaussian smoothing kernel with a time-dependent covariance matrix given by 
\begin{align}
C_t &= \Ddiff t \left(\begin{array}{cc} t^2/(3m^2) & t/(2m) \\ t/(2m) & 1 \end{array} \right).
\end{align}
The distribution of outcomes for the final position measurement is obtained by integrating the Wigner distribution over momentum: $P(x) = \int\! \mathrm{d}p \, W_T(x,p)$.

If contractive states are excluded, one can check that the Gaussian initial states producing final measurement distributions depending most sensitively on $\Ddiff$ (in the sense that the Chernoff bound exponent is maximized \cite{cover2006elements}) are always the ones for which position and momentum are initially uncorrelated; in this case, the Heisenberg uncertainty inequality is saturated and the uncertainties ($\sigma$'s) from the heuristic arguments in the main text coincide with the standard deviations of the respective Gaussian probability distributions up to factors of order unity. A similar SQL for weak forces \cite{ozawa1988measurement} can be proved with the same techniques using $\hat{H} = \hat{p}^2/2m-F \hat{x}$.

Note again that $\Fmat^a_{\phantom{a}b}$ and $\Dmat^{ab}$ were assumed to be time-independent, consistent with time-homogeneity and the Markov property, and well motivated by the observation that the time-dependent components of these parameters in real-world cases of quantum Brownian motion are often transient when sourced by thermal baths \cite{unruh1989reduction}. Care must be taken with this assumption \cite{anglin1997deconstructing,ford2005limitations}, but more complicated SQLs can be derived even when it is relaxed.  Likewise, one can straightforwardly derive traditional SQLs for time-varying forces or effective masses.

\end{document}